\documentclass[reprint,superscriptaddress,amsmath,amssymb,
aps,prl,floatfix]{revtex4-1}

\usepackage{bm}
\usepackage{float}
\usepackage{graphicx}
\usepackage[usenames,dvipsnames]{color}
\usepackage[normalem]{ulem}
\usepackage[svgnames]{xcolor}
\usepackage{bm}% bold math
\usepackage{multirow}
\usepackage{float}
\usepackage{titlesec}
\usepackage[utf8]{inputenc}

\begin{document}
\preprint{APS/123-QED}

\title{Wide critical fluctuations of the field-induced phase transition in graphite}

\author{Christophe Marcenat}
\affiliation{Univ.  Grenoble Alpes, CEA, IRIG, PHELIQS, LATEQS, F-38000 Grenoble, France}

\author{Thierry Klein}
\affiliation{Univ.  Grenoble Alpes, CNRS, Grenoble INP, Institut Néel, F-38000 Grenoble, France}

\author{David LeBoeuf}
\affiliation{Laboratoire National des Champs Magnétiques Intenses (LNCMI-EMFL), CNRS, UGA, UPS, INSA, Grenoble/Toulouse, France.}

\author{Alexandre Jaoui}
\affiliation{JEIP, USR 3573 CNRS, Coll\`ege de France, PSL Research University, 11, Place Marcelin Berthelot, 75231 Paris Cedex 05, France.}
\affiliation{Laboratoire de Physique et Etude des Mat\'eriaux (CNRS/UPMC), Ecole Sup\'erieure de Physique et de Chimie Industrielles, 10 Rue Vauquelin, 75005 Paris, France.}

\author{Gabriel Seyfarth}
\affiliation{Laboratoire National des Champs Magnétiques Intenses (LNCMI-EMFL), CNRS, UGA, UPS, INSA, Grenoble/Toulouse, France.}
\affiliation{Universit\'e Grenoble-Alpes, Grenoble, France.}

\author{ Jozef Ka\v cmar\v c\'ik}
\affiliation{Centre of Low Temperature Physics, Institute of Experimental Physics, Slovak Academy of Sciences, Watsonova 47, SK-04001 Ko\v sice, Slovakia}

\author{Yoshimitsu Kohama}
 \affiliation{The Institute of Solid State Physics, University of Tokyo, Kashiwa, Chiba 277-8581, Japan}

\author{Herv\'e Cercellier}
\affiliation{Univ.  Grenoble Alpes, CNRS, Grenoble INP, Institut N\'eel, F-38000 Grenoble, France}

\author{Herv\'e Aubin}
\affiliation{Centre de Nanosciences et de Nanotechnologies,
CNRS, Universit\'e Paris-Saclay, 91120 Palaiseau, France}
  
\author{Kamran Behnia}
\affiliation{Laboratoire de Physique et Etude des Mat\'eriaux (CNRS/UPMC), Ecole Sup\'erieure de Physique et de Chimie Industrielles, 10 Rue Vauquelin, 75005 Paris, France.}

\author{Beno\^it Fauqu\'e}
\email{benoit.fauque@espci.fr}
\affiliation{JEIP, USR 3573 CNRS, Coll\`ege de France, PSL Research University, 11, Place Marcelin Berthelot, 75231 Paris Cedex 05, France.}

\date{\today}% It is always \today, today,

\begin{abstract}

In the immediate vicinity of the critical temperature (T$_c$) of a phase transition, there are fluctuations of the order parameter, which reside beyond the mean-field approximation. Such critical fluctuations usually occur in  a very narrow temperature window in contrast to Gaussian fluctuations.  Here, we report on a study of specific heat in graphite subject to high magnetic field  when all carriers are confined in the lowest Landau levels. The observation of a BCS-like specific heat jump in both temperature and field sweeps establishes that the phase transition discovered decades ago in graphite is of the second-order. The jump is preceded by a steady field-induced enhancement of the electronic specific heat. A modest (20 percent) reduction in the amplitude of the magnetic field (from 33 T to 27 T) leads to a threefold decrease of T$_c$ and a drastic widening of the specific heat anomaly, which acquires a tail spreading to two times T$_c$. We argue that the steady departure from the mean-field BCS behavior is the consequence of an exceptionally large Ginzburg number in this dilute metal, which grows steadily as the field lowers. Our fit of the critical fluctuations indicates that they belong to the $3DXY$ universality class, similar to the case of $^4$He superfluid transition.

\end{abstract}

\maketitle
A phase transition is accompanied by sharp discontinuities of the thermodynamic properties. Quantifying entropy  by measuring the specific heat %($C$=$T\frac{dS}{dT}$) 
across the transition pins down the order of the transition and informs on the underlying microscopic interaction. Of particular interest is the critical regime of the phase transition, which allows the identification the universality class of the transition and provides direct information on the order parameter \cite{KADANOFF1967}. Critical fluctuations are important when their amplitude is comparable with the amplitude of the jump of the specific heat, $\Delta C$, which occurs roughly
when the reduced temperature $\tau=\frac{T-T_c}{T_c}$ is smaller that of others $\tau_G$ \cite{Ginzburg1960}, the Ginzburg criterion with:
\begin{equation}
\tau_G=\alpha^2 (\frac{k_B}{\Delta C\xi_m^3})^2    
\label{Ginz} 
\end{equation}
Here, $\xi_m$ is the correlation length (averaged in presence of anisotropy) and $\alpha=\frac{1}{4\sqrt{2}\pi}$ is a numerical factor. In most cases, $\tau_G \ll 1$, the critical fluctuations are located in the extreme vicinity of the transition, and therefore, hardly observable. One notorious exceptions is the $^4$He superfluid transition for which the shape of the transition is determined by the critical fluctuation \cite{Lipa96}. Near a quantum critical point, thermal fluctuations are replaced by quantum mechanical zero-point fluctuations, which can produce new quantum phases \cite{SubirBook}.

Here, we report on the electronic specific heat ($C_{el}$) of graphite, using the state of the art of calorimeters,  when all the carriers are confined in the lowest Landau levels (LLL), the so-called quantum limit, which can be easily achieved in this dilute metal. We find that this regime is marked by a steady field-induced enhancement of $C_{el}$, signaling the enhancement of electron-electron correlations. Deep in this regime, we detect a jump in $C_{el}$,  unambiguously establishing  a second-order phase transition induced by the magnetic field. As the magnetic field is reduced, the anomaly shifts to lower temperature and widens. It evolves from a BCS mean-field type transition at 33 T to a cross-over regime below 25 T. At the lowest critical temperature (T$_{c}\sim$ 1K),  fluctuations can be observed up to two times T$_c$. We identify them as critical fluctuations associated with an exceptionally large Ginzburg number due to a low energy condensation. Comparison with a number of other phase transitions shows what distinguishes this phase transition and its criticality. In graphite, the difference in the heat capacity between the normal and ordered phase within a coherence volume rapidly drops with decreasing magnetic field and becomes exceptionally low.

\begin{figure}[h!]
\centering
\includegraphics[width=\linewidth]{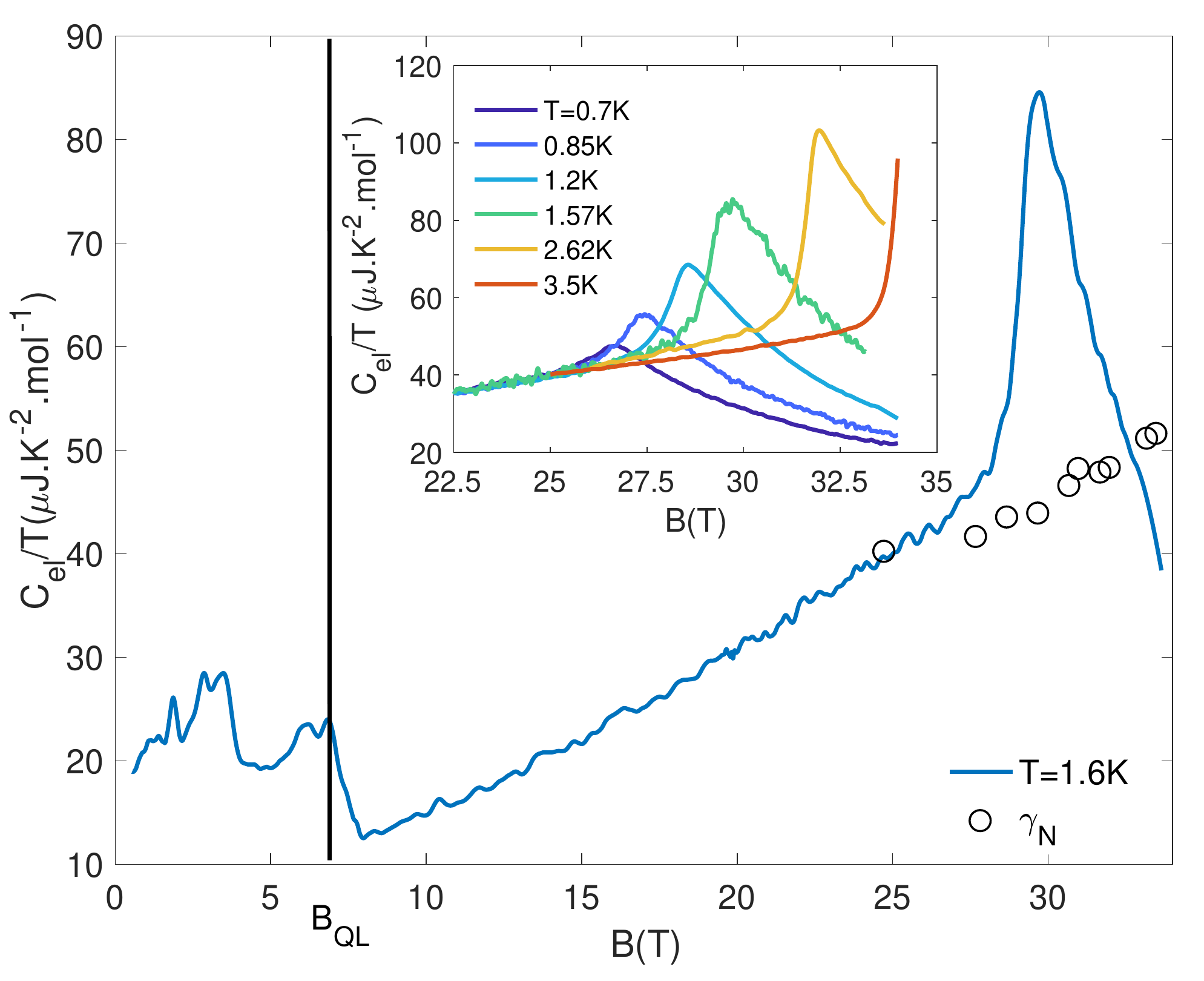}
\caption{Field dependence of $\gamma_{el}=\frac{C_{el}}{T}$ at T=1.6K.  The vertical black line indicates  the quantum limit regime. When B$> $B$_{QL}$=7.5T, holes and electrons are confined to  their  n=0 Landau levels (with two spin-polarized sub-levels for each). Open circles represents $\gamma$ in the normal state (labelled $\gamma_N$), deduced from temperature sweeps shown in Fig.\ref{FigCvsT}. The inset shows field sweeps at different temperatures.}
\label{FigCvsB}
\end{figure}

With an electronic density as low as n=4$\times10^{18}$cm$^{-3}$ and light in-plane mass carrier (m$^{\star}_{a,b}$=0.05m$_0$), a magnetic field of 7.5T, labelled B$_{QL}$, oriented along the c-axis of graphite is large enough to confine the carriers into the lowest LL (n=0). In the early 80s Tanuma and co-workers \cite{tanuma1981} discovered the onset of an electronic phase transition at B=25T and T=1.3K. Since then, extensive electrical \cite{yaguchi1998,Fauque2013b,Akiba15,Zhu2017,Arnold14,Zhu18}, thermo-electrical \cite{Fauque2011}  and ultrasound measurements \cite{LeBoeuf2017} at high magnetic field have established that graphite hosts a succession of, at least, two field induced phases \cite{Fauque2013b,Zhu18} arising from electron-hole instabilities. Depending on the nesting vector considered and the strength of the electron-electron interaction, various types of charge \cite{Yoshioka1981,Arnold14}, spin \cite{Takada1999} density waves or an excitonic insulating phase \cite{Akiba15,Zhu2017,Zhu18,Shindou2019} have been proposed. To the notable exception of ultrasound measurements \cite{LeBoeuf2017} these studies have employed transport probes. Due to the low electronic density of graphite, thermodynamic studies are challenging and, nevertheless, crucial.

\begin{figure}[h]
\centering
\includegraphics[width=\linewidth]{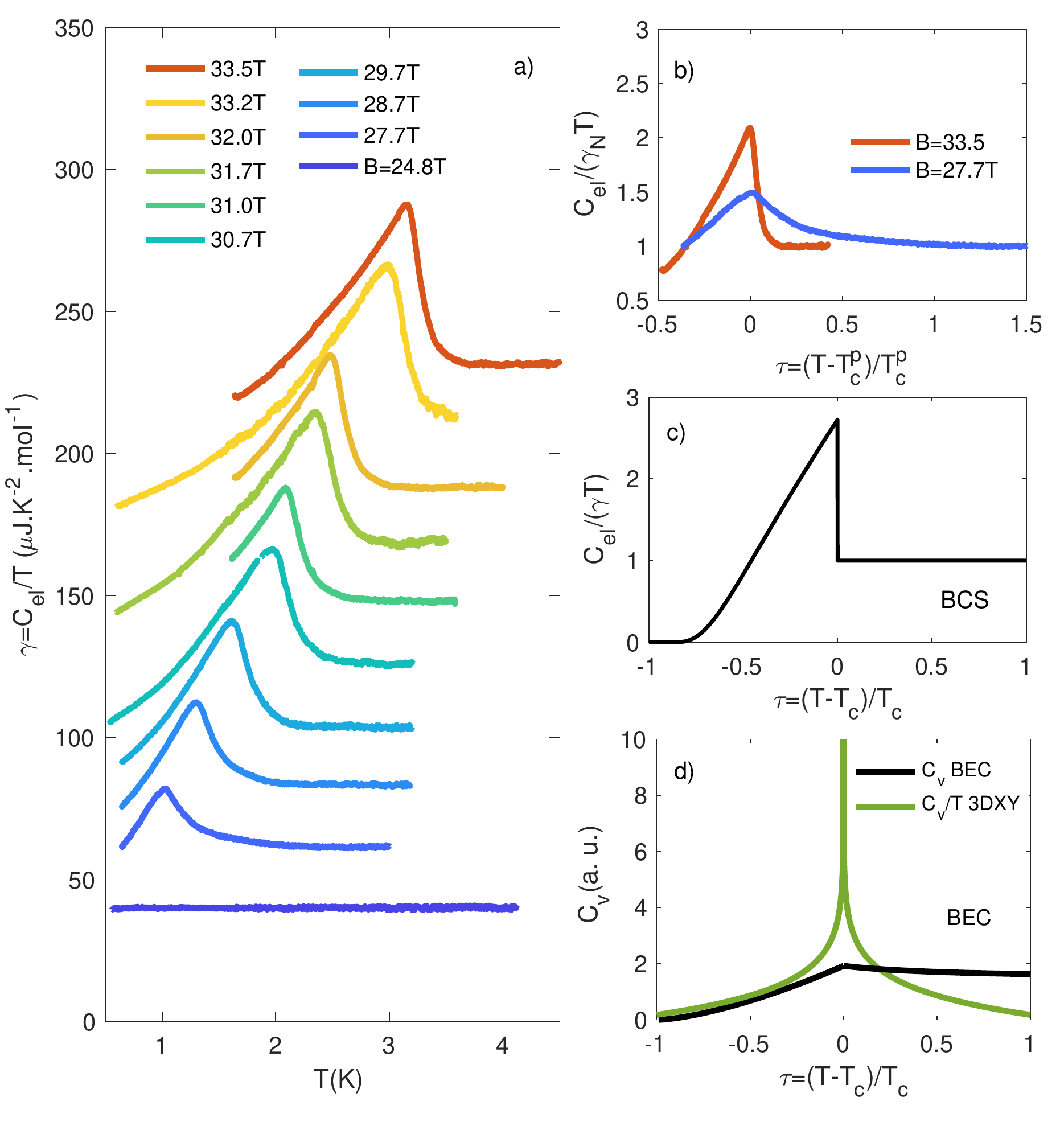}
\caption{ Temperature dependence of $\gamma$:  a) $\gamma=\frac{C_{el}}{T}$ function of temperature for different magnetic fields. b) $\frac{C_{el}/T}{\gamma_N}$ as function of the reduced temperature $\tau=\frac{T-T^p}{T^p}$ where $T^p$ is the temperature of the peak position. At the lowest field the tail of the transition extended up to two times T$_c$. c) Specific heat anomaly in a BCS transition. The amplitude of the jump at T$_c$ is such that $\frac{\Delta C_{el}}{\gamma_{el}T_c}$=1.43 d) Specific heat anomaly in a BEC transition (in black) and the singularity caused by 3DXY critical fluctuation regime (in green) in the case of the superfluid transition in helium.}
\label{FigCvsT}
\end{figure}

Fig.\ref{FigCvsB}a) shows the field dependence of $\gamma$=$\frac{C_{el}}{T}$ at T=1.6K up to 35T of graphite. The specific heat set-up is described in the supplement\cite{SM}. At zero field, $\gamma$ is as small as 20 $\pm$ 3 $\mu$J.mol$^{-1}$.K$^{-2}$ in good agreement with the value expected from the Slonczewski Weiss McClure (SWM) band model \cite{Komatsu1951,Hoeven1963}. This is several orders of magnitude smaller than in metals due to its low density and lightness of carrier. Sweeping the magnetic field, we found that $\gamma$ peaks at each LL depopulation. Above B$_{QL}$, $\gamma$ increases linearly up to 28T where it presents yet another peak, which evolves with temperature. The temperature and magnetic field dependence of the 28T peak can be tracked by field sweeps (at different temperatures), as shown in the inset of Fig.\ref{FigCvsB} and by temperature sweeps (at different magnetic fields), as shown on Fig.\ref{FigCvsT} a). 

Let us begin by commenting the evolution of $\gamma$ with magnetic field  between  10 T and  35 T. Its magnitude in the normal state (labelled $\gamma_{N,}$) can be deduced from temperature sweeps, shown in Fig.\ref{FigCvsT}a), and is represented by open black circles in Fig.\ref{FigCvsB}.  $\gamma_{N}$ increases linearly with magnetic field. The threefold enhancement between 10 T and  35 T is  larger than the enhancement of $\gamma$ reported in Sr$_3$Ru$_2$O$_7$ across its quantum critical point \cite{Rost16549}. This remarkable enhancement is driven by the change of the density of states (DOS) induced by the magnetic field. When B$>$B$_{QL}$ the DOS of the LLL is the product of the in-plane degeneracy (which scales linearly with the field) and the one dimensional DOS along the field direction. As long as the Fermi energy (E$_F$) is far from the bottom of the LLL $\gamma_{N,el}$ scales linearly with B and with $\frac{\partial\gamma}{\partial B} \propto\sqrt(\frac{m^{\star}_{z}}{E_F})$  where $m^{\star}_{z}$ is the mass along the magnetic field and E$_F$ (see the supplement \cite{SM}). The small E$_F$$\simeq$40meV and the large c-axis mass m$^{\star}_{c}$$\simeq$10-20$m_0$ of graphite set the large slope, which we find to be $\frac{\partial\gamma_{N,el}}{\partial B}$=2.2 $\pm$ 0.5$\mu$J.K$^{-2}$.mol$^{-1}$.T$^{-1}$). Decades ago, starting from the SWM model, Jay-Gerin computed $\frac{\partial\gamma}{\partial B}$ =1.8$\mu$J.K$^{-2}$.mol$^{-1}$.T$^{-1}$ \cite{Jay1976} in quantitative agreement with our result.

\begin{figure}[h]
\centering
\includegraphics[width=\linewidth]{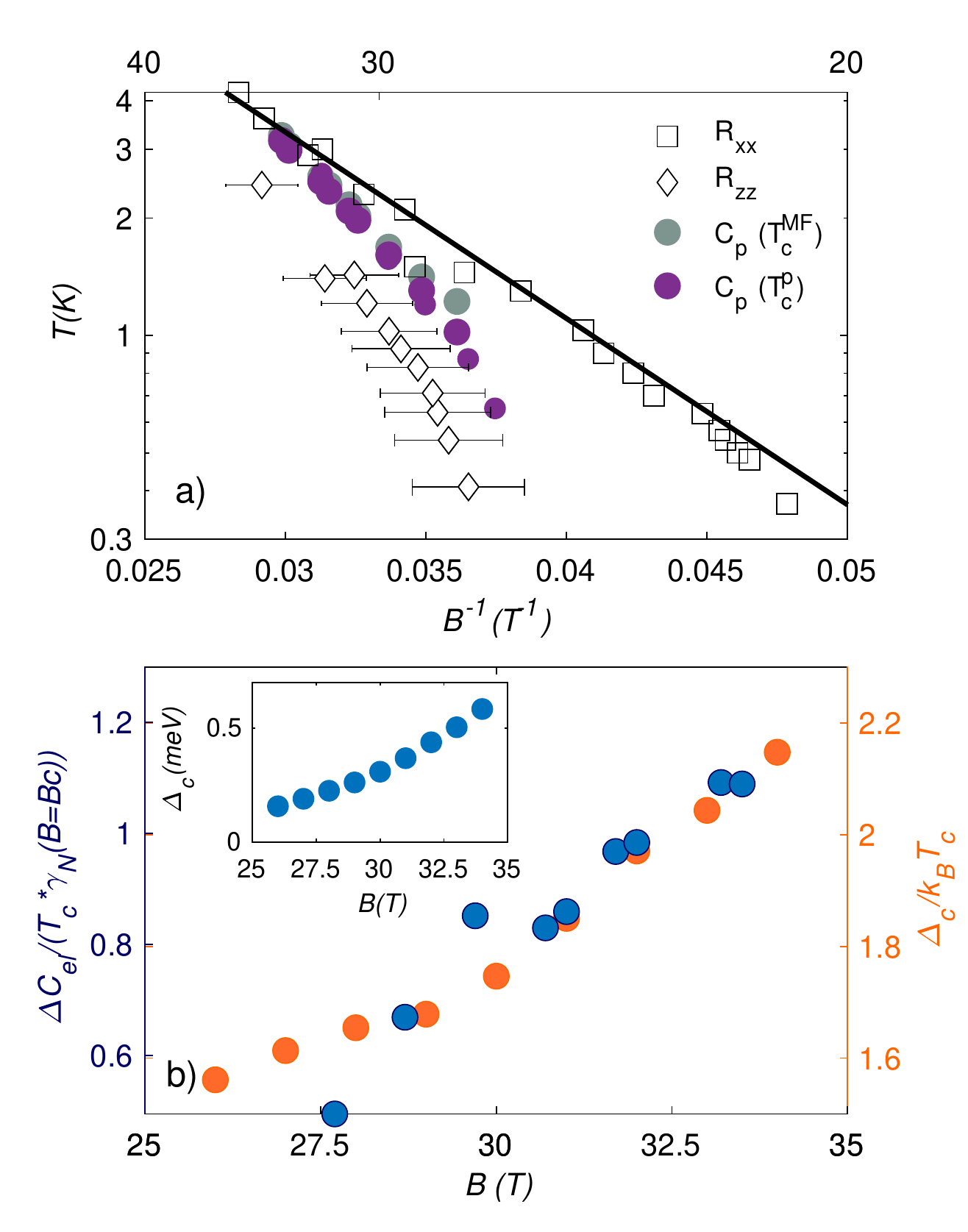}
\caption{a) Phase diagram of graphite showing the evolution of critical temperature (T$_c$) as a function of the inverse of magnetic field, (T$_{c}$,B$_c^{-1}$), according to different experimental probes. Anomalies in R$_{xx}$ (black open square points) and R$_{zz}$ (black open diamonds points) (from \cite{LeBoeuf2017}) are compared with peaks in $C_{el}$ (T$^{p}_c$ in purple close circles). The black line represents T$_{c}$=T$^{\star}\exp{\frac{-B^{\star}}{B_{c}}}$ and reasonably describes the evolution of $R_{xx}$ anomalies. b) Field dependence of $\frac{\Delta C_{el}}{\gamma_{N,el}T_c}$ and $\frac{\Delta_{c}}{k_{B}T_{c}}$. $\Delta_c$ is the activation gap deduced from the c-axis resistance measurements (see \cite{SM} for the raw datas). The field dependence of $\Delta_c$ is shown in the inset.}
\label{FigDP}
\end{figure}

At the highest magnetic field (B=33.5T), the specific heat anomaly occurs at T$_c\simeq$3K (see  Fig.\ref{FigCvsT}a)). The transition shifts to lower T$_c$ with decreasing magnetic field, in contrast to a superconducting transition. By reducing the magnetic field by twenty percent, i.e. by passing from 33.5T to 27.7T, T$_c$ decreases by a factor of 3. Fig.\ref{FigDP}a) shows how T$_c$ derived from the peak in $\frac{C_{el}}{T}$ compares with those deduced from anomalies in the in-plane (R$_{xx}$) and the out-of-plane resistance (R$_{zz}$) measurements. Each symbol represents a T$_c$ (B) (or B$_c$(T)) anomaly. The vertical axis represents $\ln(T)$  and the horizontal axis B$^{-1}$. Thus, the BCS-like expression T$_{c}$=T$^{\star}\exp{\frac{-B^{\star}}{B_{c}}}$ \cite{Yoshioka1981} becomes a straight line, which is the behavior of T$_c$(B) according to R$_{xx}$ measurements. In this formula, $T^{\star}$ and  B$^{\star}$ are phenomenological temperature and field scales and the underlying assumption is that the DOS linearly increases with magnetic field \cite{Yoshioka1981}. The onset of ordering has different manifestations in R$_{xx}$ (which jumps by 30$\%$ but does not diverge) and in R$_{zz}$ \cite{Fauque2013b,Zhu18}, which shows an activation behavior and a detectable energy gap (see \cite{SM} for complementary measurements). As seen in Fig.\ref{FigDP}a), the peak in $\frac{C_{el}}{T}$ tracks the onset of the activation energy in R$_{zz}$ and below 3 K, they both occur below the R$_{xx}$ anomaly and the BCS line.  Fig.\ref{FigDP}b), shows the evolution of the ratio of the activation gap, $\Delta_c$, to the critical temperature with magnetic field, which is close to what is expected in the BCS picture and it weakens  with decreasing field. The  same figure shows that the normalized specific heat jump  $\frac{\Delta C_{el}}{\gamma T_c}$ presents a similar evolution towards weak coupling as the field decreases. 

The steady evolution towards weak coupling with decreasing magnetic field is accompanied by a drastic change in the shape of the specific heat anomaly (see Fig.\ref{FigCvsT}a and b). At B=33.5T, the amplitude of the jump is $\frac{\Delta C_{el}}{\gamma_NT_c}$=1.1, smaller than what is expected for a mean-field weak-coupling case and has a small tail caused by fluctuations above the critical temperature. With decreasing T$_c$ the transition widens, $\frac{\Delta C_{el}}{\gamma T_c}$ decreases and the tail extends to higher temperatures. At B$_c$=27.7T, the tail of the transition extends up to twice T$_c$. The anomalies at two fields differing merely by a factor of 1.2 are compared in Fig.\ref{FigCvsT}. At B=33.5T, the anomaly looks similar to a  mean-field BCS transition (sketched in Fig.\ref{FigCvsT}c)), but at  B$_c$=27.7T, it acquires a cusp shape and the fluctuating contribution weighs as much as the mean-field jump. This is to be compared and contrasted with the case of the superfluid transition in $^4$He (sketched in Fig.\ref{FigCvsT}d)). In the latter case, there is no specific heat jump \cite{feynman1953} and the 3DXY critical fluctuations induce the non-analytical behavior of the specific heat across the '$\lambda$'-transition \cite{Lipa96}. Thus, what distinguishes the case of the field induce state of graphite from a typical superconductor and the archetypal superfluid is the simultaneous presence of a BCS jump and strong fluctuations and the contrasting evolution of these two components of the specific heat anomaly in a narrow field range. 

\begin{figure*}
\centering
\makebox{\includegraphics[width=1\textwidth]{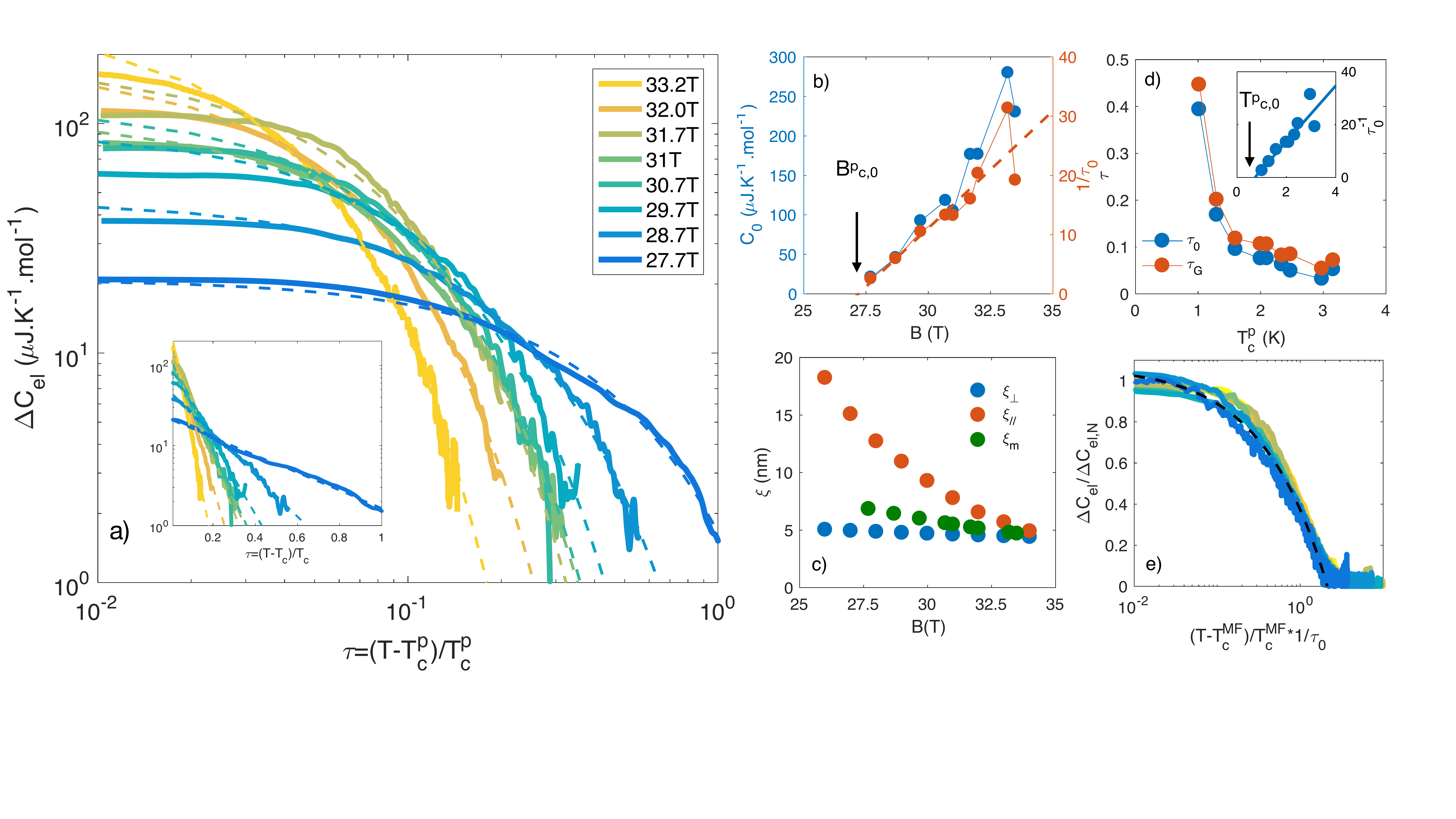}}
    \caption{Fluctuations regime of the electronic specific heat: a) $\Delta C_{el}=C_{el}-\gamma_{N}T$ as function of $\tau$=$\frac{T-T_c}{T_c}$ for $\tau>0$. The dot lines correspond to a phenomenological fit $\Delta C_{el}=C_0\exp(-\frac{\tau}{\tau_0})$. Inset shows $\ln(\Delta C_{el})$ as function of $\tau$. b) Field dependence of $C_0$ and $\tau^{-1}_0$. The red dot line is linear fit of $\tau^{-1}_0$. c) Field dependence of $\xi_{\perp}=\ell_B$, $\xi_{//}$=$\frac{\hbar v_{F,//}}{\pi\Delta_c}$ where $v_{F,//}$ is the Fermi velocity along the field, $\Delta_c$ is the gap deduced from the c-axis resistance (shown on Fig.\ref{FigDP}b)) and $\xi_m$=$(\xi^2_{\perp}\xi_{//})^\frac{1}{3}$. d) Temperature dependence of $\tau_0$ compare with the Ginzburg criterion $\tau_G$. Inset shows a plot $\tau^{-1}_0$ vs $T^p_c$ with a linear fit (blue line).  e) Normalised plot $\Delta C_{el}$ as function of $\frac{(T-T^{ren}_c)}{T^{ren}_c\tau_0}$ where $T^{ren}_c$=$T^p_c(1+0.5\tau_0)$. The black dot lines is fit with the 3DXY model (see the text and \cite{SM}).}
   \label{FigFluc}
\end{figure*}

Fig.\ref{FigFluc}a) shows $\Delta C_{el}$=$C_{el}-\gamma T$ as a function of the reduced temperature $\tau=\frac{T-T^p_c}{T^p_c}$. Here $T^p_c$ is the temperature at which specific heat peaks. $\Delta C_{el}$ quantifies the excess in specific heat relative to a mean-field transition. The log-log plot clearly shows that it does not evolve as power law and points therefore to a non-Gaussian origin (for which $\Delta C_{el}^{Gauss.}=\frac{1}{\xi^{D}_0\tau^{2-D/2}}$ with $D$ is the dimension and $\xi_0$ is the BCS coherence length \cite{LarkinBook}). On the other hand, as shown in the inset of the same figure, a simple empirical law of the form : $\Delta C_{el}= C_0 \exp(-\frac{\tau}{\tau_0})$ captures the evolution of the data with B and $\tau$. The field dependence of the two parameters $\delta C_0$ and $\tau_0$ is shown in Fig.\ref{FigFluc}b). With decreasing magnetic field  $\tau_0$ increases and its extrapolation implies $\tau_0^{-1}=0$  when B$<$27 T, where no specific heat anomaly is detected down to 0.6K ( see Fig.\ref{FigCvsT}a). As seen in  Fig.\ref{FigFluc}e), it is possible to collapse all the curves on top of each other by plotting $\Delta C_{el}/C_0$ \textit{vs.} $\frac{(T-T^{ren}_c)}{T^{ren}_c\tau_0}$ where $T^{ren}_c$ is a renormalised $T^{p}_c$ ( and taken as $T^{ren}_c$=$T^{p}_c(1+0.5\tau_0)$).

The success of this simple scaling procedure has implications for the origin of the broadening of the transition caused by a small variation in the amplitude of the magnetic field. It is unlikely that disorder plays a major role. In this layered material, however, the in-plane and out-of-plane length scales differ by several orders of magnitude. The mean free path within the graphene planes, $\ell^{\perp}_e$ is very long, in the range of 5$\mu$m to 50$\mu$m,  three orders of magnitude longer than $\frac{1}{k_{F,\perp}^{-1}}$=7nm, and the magnetic length,  $\ell_B$=$\sqrt(\frac{\hbar}{eB})$=5nm at 25 T which is a plausible candidate to represent the in-plane coherence length ($\xi_{\perp}$) beyond the quantum limit. On the other hand, the mean free path along the c-axis, $\ell^{//}_e$ is two or three orders of magnitude shorter than $\ell^{\perp}_e$. However, it remains longer than the c-axis inter-electron distance, d$_{e-e,//}\simeq\frac{1}{k_{F,//}}\simeq1nm$ and exceeds the c-axis coherence length ($\xi_{//}$). The latter can be estimated using the BCS coherence length $\xi_{//}$=$\xi_{0}$=$\frac{\hbar^2 k_{F,//}}{\pi m^{\star}_z\Delta_c}$. Here $\Delta_c$ is the c-axis gap (deduced from the activated behavior of R$_{zz}$ and shown in the inset of Fig.\ref{FigDP}b)) and $k_{F,//}$=$\frac{\pi}{4a}$ (where $a$ is the interlayer distance). As seen in Fig.\ref{FlucSM}d) $\xi_{//} (33T) \simeq$5nm and $\xi_{//} (25 T) \simeq$ 15nm. The latter number may approach the estimated $\ell^{//}_e$ below 25T. Therefore the role played by stacking disorder across the planes in causing the broadening cannot be rule out at the lowest magnetic field.

The most plausible source of the observed broadening are critical fluctuations. This can be seen by quantifying the Ginzburg criterion and the expected width of the thermal window for critical fluctuations. Plugging the measured $\Delta C_{el}$ and the estimated $\xi_{//,\perp}$ ($\xi_{m}=(\xi^2_{//}\xi_{\perp})^\frac{1}{3}$) in Eq.\ref{Ginz}, we deduced $\tau_G$. As seen in Fig.\ref{FigFluc}d), this estimation of $\tau_G$ closely matches $\tau_0$ deduced from our fits to the data. At the highest field, B=33T $\tau_G\simeq0.05$, which is already non negligible. As the field decreases to 27.7T, $\tau_G$ becomes as large as 0.45, which is exceptionally large in comparison to any other known second-order phase transition. This is seen in Table \ref{TabtauG} which compares our case with a few other systems. The list consists of a conventional superconductor (Sn), two superconductors with higher T$_c$s and shorter coherence lengths (MgB$_2$ and YBa$_2$CuO$_7$), and a Charge Density Wave (CDW) solid (K$_{0.3}$MoO$_3$). One can see that it is the small magnitude of $\Delta C_{el}$, which distinguishes graphite, a system in which 10000 atoms share a single electron and hole. Even at 33 T, the amplitude of the specific heat jump is many orders of magnitude smaller than others. This $\Delta C_{el}\simeq$0.16mJK$^{-1}$mol$^{-1}$ (combined with a molar volume of V$_{m}$=5.27 cm$^{3}$.mol$^{-1}$) implies that the difference in average specific heat of the normal and ordered phases within a coherence volume becomes comparable to the Boltzmann constant ($\Delta C\xi^3_m\simeq 0.3k_B$), making this competition critically fragile. With decreasing magnetic field, fluctuations grow and the transition widens because of further decrease in $\Delta C_{el}$.

\begin{table}[htbp]
 \caption{\label{tab:table2} Comparison of $\tau_G=\frac{1}{32\pi^2}(\frac{k_B}{\Delta C\xi_{0}^3})^2$ for different superconductors (SC) (Sn, MgB$_2$ and YBa$_2$CuO$_7$), the CDW system (K$_{0.3}$MoO$_3$) and in graphite. $\Delta C$=$\Delta C_{el}\times V_m$ is the amplitude of the jump of specific heat at the transition in J.K$^{-1}$.m$^{-3}$, V$_m$ is the molar volume, $\xi_0$ the BCS coherence length. For anisotropic compounds the average value of $\xi_0$, $\xi_m$, is used. }
 \begin{ruledtabular}
\begin{tabular}{lclccccl}
Sample & $\Delta C$(J.K$^{-1}$.m$^{-3}$)& $\xi_0$($\AA$) & $\tau_G$\\
 \hline
Sn (SC)\cite{KADANOFF1967}  & 800 & 2300 & $10^{-14}$\\
MgB$_2$ (SC)\cite{Park2002}  & 3360 & 26 & $10^{-3}$  \\
K$_{0.3}$MoO$_3$ (CDW)\cite{Brill1995,McKenzie1995} & 35800 & 6.2($\xi_{m}$) & 0.01 \\
YBa$_2$CuO$_7$ (SC) \cite{Junod2002} & 4993 & 6.7($\xi_{m}$) & 0.2 & \\
graphite (33.2T) & 30.1 & 47($\xi_{m}$) & 0.05  \\
graphite (27.7 T) & 3.7 & 68($\xi_{m}$) & 0.45 \\
\end{tabular}
\end{ruledtabular}
\label{TabtauG}
\end{table}

So far we have discussed the temperature dependence of $\Delta C_{el}$ through a phenomenological two-parameter exponential fit. As shown in Fig.\ref{FigFluc}e), on the whole temperature range, it is equally possible to fit $\Delta C_{el}$ with a simplified version of the asymptotic form of 3DXY universality class expression with three parameters : $A_0(1+C_0\lvert{\tau}^{0.5}\rvert+D_0\tau)$ (see the supplement \cite{SM}). This universality class provides a natural explanation for the saturation of $\Delta C_{el}$ at low $\tau$ due to its  almost vanishing critical exponent ($\alpha\approx$-0.01\cite{Guillou1985}). Therefore, we conclude that the observed broadening of the transition is caused by  critical fluctuations belonging to the 3DXY model. Let us note that Gaussian fluctuations expected outside the critical window, i.e. for $\tau>\tau_G$ would bring a correction to the mean field behavior. Such a correction is not detected in our data, presumably because of the width and predominance of the critical fluctuations.

Let us also note that below a critical temperature,  T$^{p,0}_{c}\simeq$1K, and a critical field  B$^{p,0}_c\simeq$ 25T the entrance in the ordered phase is no more accompanied by a well defined jump in $C_{el}$ but it is replaced by a cross-over. In this case, like in the case of the $\lambda$-transition, critical fluctuations dominate the transition. However, in contrast to the $\lambda$-transition \cite{Lipa96}, they are not restricted to the extreme vicinity of the transition.

The present result helps to understand the evolution of the observed anomalies in the Nernst effect \cite{Fauque2011,Zhu2020} (a measure of the entropy per carrier \cite{Bergman2010}) and the ultra-sound measurements \cite{LeBoeuf2017}  caused by the transition. At low temperature, one expects that ordering induces a smooth variation in entropy (see \cite{SM}) and therefore a rounded drop in the Nernst response \cite{Fauque2011,Zhu2020}. As the temperature increases the  mean field component of the transition strengthens and  the Nernst anomaly becomes a clear kink. Furthermore, the origin of the relatively large jump in the sound velocity becomes clear. It was argued \cite{LeBoeuf2017} that it can be caused by either a strong anisotropy in the strain dependence of T$_c$ or a large jump in  the specific heat (for example due to a putative lattice deformation accompanying the transition). According to the present study, $\Delta C\approx \gamma T_{c}$ and the transition is purely electronic. Therefore, we can safely conclude that the jump in the sound velocity is caused by a strong anisotropic strain dependence of T$_c$.

Future studies of the specific heat at higher magnetic field would shed light to the BCS-BEC crossover as one approaches the maximum transition temperature around 47T, where the degeneracy temperature and critical temperature become close to each other\cite{Zhu2020}. Specific heat studies on other dilutes metals pushed to extreme quantum limit and hosting field-induced state (such as bismuth \cite{Zhu2017Bi,Iwasa2019}, InAs \cite{alex2020giant}, TaAs \cite{Ramshaw2018} or ZrTe$_5$ \cite{Tang2019}) could bring interesting insights.

In summary, we measured the specific heat of graphite in high magnetic field and detected a second-order phase transition jump at high field. The specific heat anomaly drastically evolves in a narrow field window as a consequence of the change in the balance between critical fluctuations and a mean-field jump. The field-induced phase transition in graphite emerges from this study as possessing an exceptionally wide critical window compared to any other electronic phase transition.

\begin{acknowledgments}
We thank Zengwei Zhu for useful discussions. We acknowledge the support of the LNCMI-CNRS, member of the European Magnetic Field Laboratory (EMFL). This work was supported by JEIP-Coll\`{e}ge de France and by the Agence Nationale de la Recherche  (ANR-18-CE92-0020-01; ANR-19-CE30-0014-04). J.K. was supported by the Slovak grants No. APVV-16-0372 and VEGA 2/0058/20.
\end{acknowledgments}
\bibliography{apssamp}

\end{document}